# Forecasting solar radiation during dust storms using deep learning

B. Ravindra

*Abstract*—Dust storms are common in arid zones on the earth and others planets such as Mars. The impact of dust storms on solar radiation has significant implications for solar power plants and autonomous vehicles powered by solar panels. This paper deals with the analysis of solar radiation and power output of a rooftop photovoltaic plant during a dust storm and proposes a forecasting methodology using deep learning network. The increased aerosol content due to dust storms increases the diffuse component of the solar radiation. This effect persists for a long duration and can impact the quality of forecasting of solar radiation. Deep learning networks that capture long range structure can improve the quality of solar radiation forecasting during dust storms. These results can help explain the sudden drop in power output of solar plants due to dust storms originating in another continent. They can shed light on mysterious cleaning events in autonomous vehicles powered by solar panels to be used in space missions.

*Index Terms*—Solar radiation, Dust storms, Solar photovoltaic power plants, Deep learning. Mars rovers

## I. INTRODUCTION

FORECASTING of time series data has assumed greater importance in a number of disciplines such as meteorology, nonlinear and chaotic dynamics and renewable energy during the last five decades. Since the work of Lorenz it is widely accepted that long range weather prediction is impossible due to its inherent chaotic nature. Several researchers have studied short term forecasting of chaotic time series using techniques such as time delay embedding and neural networks [1-3]. Some of these techniques have been applied to numerical and experimental data obtained from classic examples of systems exhibiting chaos. Time series data from several domains such as weather prediction, solar radiation forecasting, and economics often involve complex long range interactions [3-5]. Recent discoveries in machine learning have helped improve quality of predictions of complex time series data. Conventional recurrent neural networks mostly use data from recent observations [6]. This limits their applicability when time series with long-range structure needs to be analyzed. This forgetfulness may lead to instability when predictions are carried out. Networks with a longer memory may improve the stability and enhance the quality of predictions. Long Short-term Memory (LSTM) is an example of this category of neural networks that has been found to improve predictions in a number of applications such as speech recognition and robotics [7-10]. This article deals with one such case of solar radiation prediction due to dust storms and explores implications for solar powered autonomous vehicles and solar power plants. The underlying physics of solar radiation during and after a dust storm shows long range interactions that can be captured by LSTM networks.

The impact of dust storms on people living in the arid and semi-arid zones has received wide attention. Most of these investigations have studied health and agricultural and transportation aspects. It is known that the dust storms from Sahara desert can impact cities in Europe and North America and sand storms in Middle East can continue towards south Asia [11-12]. The impact of these events on the solar power generation needs detailed investigation.

The physics of dust aerosol emission, the weather phenomena that trigger dust storms have been reviewed by J.F. Kok et al [13-15]. The nature of hysteresis phenomenon involved in Martian dust storms has been discussed by J.F. Kok [14]. An interesting scaling theory for the size distribution of emitted dust aerosols and its implication for impact on Earth's radiation budget and in understanding past and predicting future climate changes have been studied by J.F. Kok [15]. It may be noted that the implications of solar dimming and brightening phenomenon for solar photovoltaic (PV) power plants have also been examined recently by climate scientists [16]. This study deals with long term averages and trends but the case of short term forecasting is not considered. The impact of climactic events such as sand storms is very significant and changes in solar radiation data during these events impact solar resource assessment and forecasting. Solar panel selection in autonomous vehicles such as Mars rover needs to consider events such as dust storms. The case of mysterious cleaning events of solar panels in Mars exploration in enhancing the life of the rover has been discussed since 2004[17]. It is interesting to provide quantitative estimation/forecasting of how solar radiation varies during such dust storm events for future space missions. In the case of terrestrial sand storms, meteorological weather stations have the capability to monitor solar radiation and other weather related parameters to monitor sand storms. A case of one such sand storm that emanated from Middle East and

This paper is submitted on 21$^{st}$ August 2018.
B. Ravindra is with the Department of Mechanical Engineering, Indian Institute of Technology, Jodhpur, Nagaur Road, Karwar, Rajasthan, 342 037 INDIA (e-mail: ravib@ iitj.ac.in).

spread to East Asia is considered in this article. Detection of these events based on observations from microwave satellites, visible and infrared instruments as well as TOMS Aerosol index is presented in [11]. These results are also correlated with results obtained from sky radiometer at the ground station. Here this sand storm event and its impact on forecasting of solar radiation and performance of solar PV panels are discussed in detail. These results can also be extended to include other regions where similar events occur [11-12,18-19].

II. SOLAR RADIATION DURING THE DUST STORM

Terrestrial dust storms have a spatio-temporal aspect as they move from one continent to the other. Satellite data is often useful in tracking the spatial movement. It is the sky radiometer data and solar radiation data from pyranometers at ground station which give valuable information on temporal evolution of dust storm at a given site. The description of one such storm is given in [11]. On March 20, 2012, a "super sandstorm" originated in the Arabian Sea which reached as far as Southeast Asia engulfing all the regions in between including India. This dust storm persisted till $23^{rd}$ March 2012 [11]. The city of Jodhpur (Latitude $26.2^{0}$N and Longitude $73^{0}$E and 290m above MSL) has witnessed this storm and The Indian Meteorological Department (IMD) weather station at Jodhpur has monitored solar radiation and other weather related parameters during the sand storm. It is interesting to investigate the effect of aerosols on the diffuse component of solar radiation during and after the dust storm. Towards this end, the hourly averaged values of diffuse component of solar radiation during these days: One day before the dust storm ($18^{th}$ march 2012), during the dust storm ($19^{th}$ March 2012) and three days after the dust storm ($20^{th}$ to $23^{rd}$ March 2012) is shown in Fig 1. It can be observed that due to increased aerosol content during the dust storm, the diffuse horizontal irradiance (DHI) value increased substantially on $20^{th}$ March 2012 (250% increase in peak value from previous day). It continued to be high even after the dust storm. It is interesting to see the long range structure in this event. The sand storm continues to have an effect on the diffuse component of solar radiation at the given site and this has implications for solar radiation forecasting to be explored in the next section. A solar PV crystalline silicon cell radiation sensor also provided valuable data for understanding the storm event. The solar cell sensor inclined at the latitude of the site provided radiation data as shown in Figure 2 for a ten day period during this event. The data is from 6.45 AM to 7.00 PM at 15 minute interval. It can be seen that on the day of the storm the radiation value dropped significantly and continued to have an effect for the subsequent week. It can be seen that the drop in the peak value is around 21% on the day of the sand storm. Thus, in the absence of measured values of DHI and global horizontal irradiance, GHI (for which a shaded disc pyranometer and a pyranometer are required respectively), a cell radiation sensor may suffice.

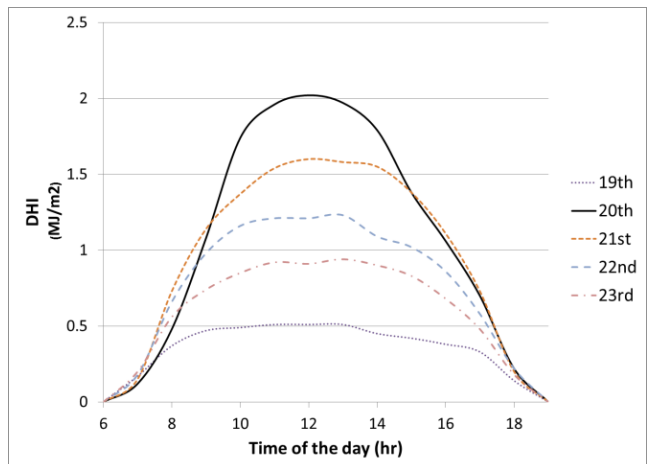

Fig 1. DHI (MJ/m$^2$) during $19^{th}$ to $23^{rd}$ March 2012

This is a cheaper alternative though it may not capture the entire spectral content of incident solar radiation.

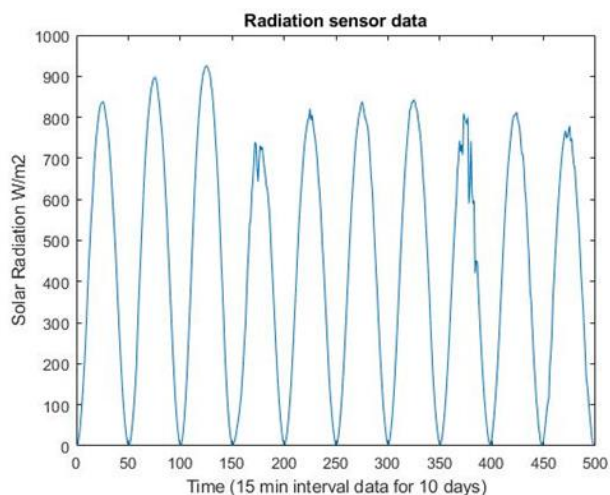

Fig 2. Radiation data during $17^{th}$ to $26^{th}$ March 2012

III. SOLAR RADIATION Forecasting DURING THE EVENT

The long range effect in the time series data due to sand storm needs to be considered while developing solar radiation forecasting algorithms. Towards this end, a Long Short-term Memory (LSTM) network is used. This network was first presented by Hochreiter and J. Schmidhuber in 1997. A good review of LSTM networks can be found in references [7-10]. Applications of LSTM networks to a number of disciplines such as chaotic dynamics, hand writing recognition, speech recognition, robotics, smart grid, self-driving cars can be found in references [7-10]. The standard recurrent neural networks (RNN) have vanishing gradient problem while training, may have instability and also under predict the solar radiation. LSTM networks have been shown to perform well in a number of applications with long range interactions. The application of this network for forecasting solar radiation during and after sand storm is considered here. The network is trained on the first 90% of the time series (9 days) and tested on the last 10% (last day). The LSTM layer has 200 hidden units. Typical parameters of the network are as follows: MaxEpochs=250;GradientThreshold=1;InitialLearnRate=0.00

5;LearnRateDropPeriod=125; LearnRateDropFactor=0.2; The time series with the forecast on tenth day is shown in Figure 3. It can be seen that the maximum solar radiation predicted is higher than the actual (compare Figure 2 and 3).

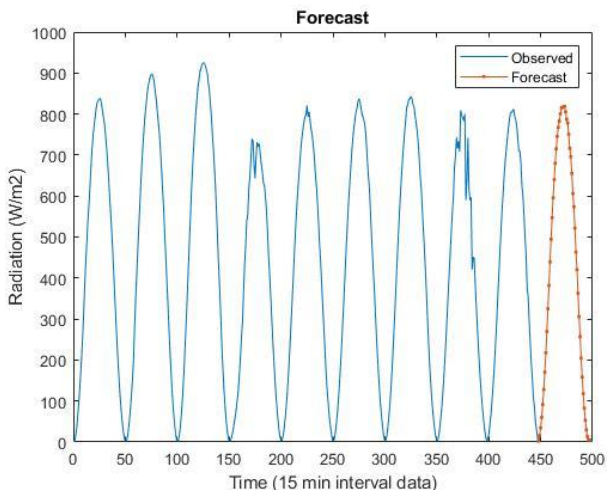

Fig 3. Radiation forecasting during 17[th] to 26[th] March 2012

The exploded view of the forecast for 26[th] March 2012 is shown in Figure 4.

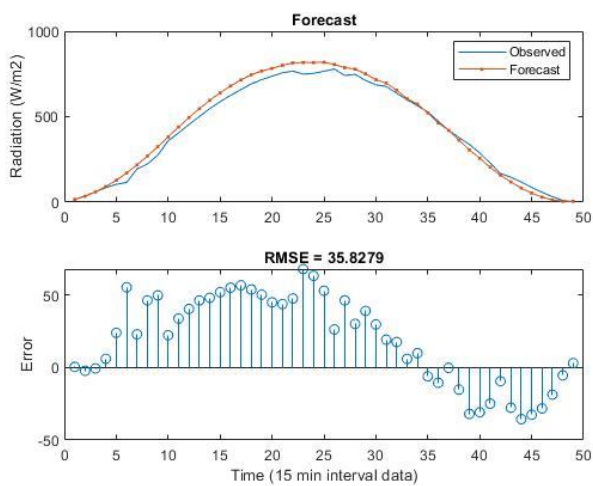

Fig 4. Radiation forecasting error on 26[th] March 2012

It may be seen that LSTM network over predicts for most of the time during the day as seen by the RMSE error shown in figure 4. This is in contrast to linear forecasting algorithms which are known to under predict solar radiation. The result can be improved upon by using the updated values as shown in Figure 5. LSTM networks are known to perform well when large amounts of data are available. Thus it is expected with a higher sampling rate, the network would have performed better. It may be remembered that inherent limitations in the PV cell radiation sensor data have to be kept in mind while judging the RMSE error. Soiling errors are common in radiation measuring instruments. But the cell sensors are perhaps less sensitive to these errors than the standard pyranometers used to measure the global horizontal irradiance. The solar radiation value thus predicted can be used to compute expected power generation as explained in the next section.

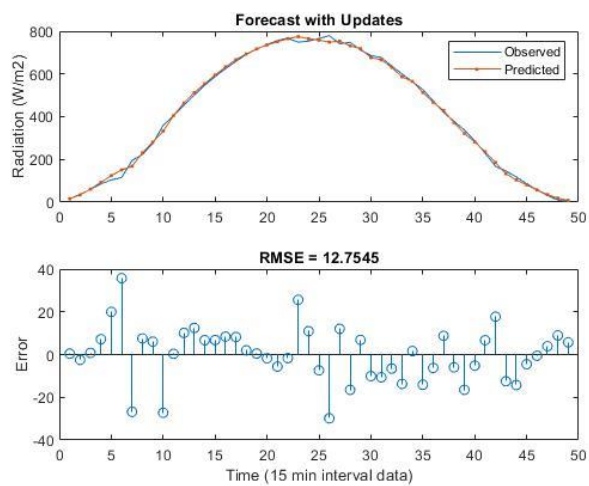

Fig 5. Solar Forecasting with updates on 26[th] March 2012

IV. SOLAR PANEL POWER OUTPUT DURING DUST STORM

It is instructive to examine the maximum power output from an existing rooftop PV plant at the site (which is used as a renewable energy laboratory for graduate students) and correlate the effect of dust storm in that context. The measured AC power from PV panels installed at the site for the ten days of the storm are shown in Figure 6. It is seen that the after a drop in power output on the day of the sand storm, the increase in power on the next day after the storm is significant though the radiation data is not as high as that on the previous day of the storm. This is akin to "mysterious cleaning events" that was coined in the case of dust storms observed on mars rover. It is not clear whether the drop in temperature due to dust storm has caused the rise of power output after the storm or the cleaning of the accumulated dust due to the increased wind speed of the storm. Comparing Figures 2 and 6 one can say that the peak power output of Figure 6 (unlike the radiation value of Figure 2) decreases monotonously after 20[th] March 2012. These results can be explained qualitatively by using a simple equivalent circuit model [20]. With this model the maximum power output from a PV cell ($P_{max}$) at a given cell temperature $T$ [20] is given by the following equation:

$$P_{max} = \frac{\frac{eV_{mp}^2}{kT}}{\left(1 + \frac{eV_{mp}}{kT}\right)}(I_{sc} + I_0)$$

Where, $e$ is the charge of an electron, $k$ is the Boltzmann constant, $T$ is the absolute temperature of the cell ($^0$K), $I_{sc}$ is the short circuit current and $V_{mp}$ is the voltage at which the power is maximum. $I_0$ is the reverse saturation current which can be calculated as shown in Ref [20]. The changes in short circuit current can be considered as proportional to changes in solar radiation. Hence a decrease in solar radiation due to the dust storm results in a proportional decrease in short circuit current which in turn causes the power from the cell to drop. It is known that a dust storm event is often accompanied by a drop in ambient temperature and increase in wind speed. These factors contribute to a decrease in cell temperature and cleaning of the panel (if the dust particles do not stick to the panels). Given temperature coefficients of the PV module, it is possible to quantitatively examine the effect of this phenomenon. There are several empirical models to predict the module back surface temperature [21]. The measured temperature on the back side of the solar panel is shown in Figure 7. This figure confirms the role of temperature in the rise of power output after the storm. Thus, the net drop or rise in the power output due to dust storms is an interplay of solar radiation, wind speed and cell temperature. Examining the radiation data alone cannot explain the cleaning events or the drop in power output due to dust storms.

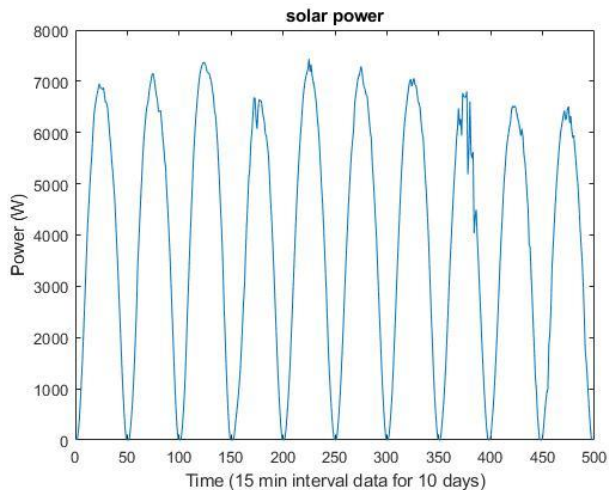

Fig 6. Power output during 17$^{th}$ to 26$^{th}$ March 2012

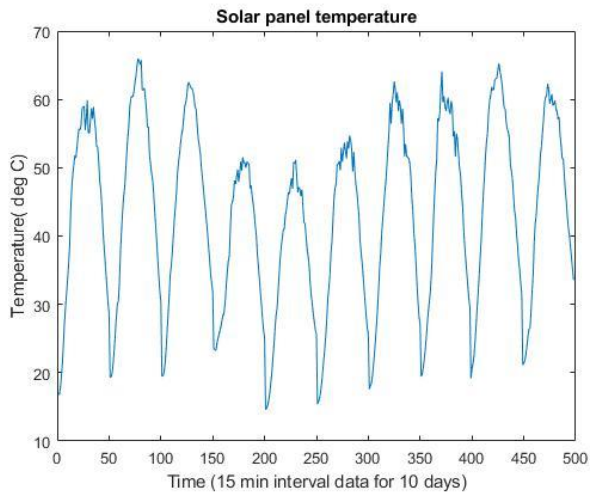

Fig 7. Module back face temperature variation

## V. CONCLUSIONS

Dust storms on the earth and other planets strongly influence the solar radiation needed to generate power through photovoltaic panels. The long range structure caused by aerosols in the diffuse solar radiation during dust storms can be incorporated in forecasting algorithms by using LSTM networks. It is interesting to see that the mysterious cleaning events in the case of Mars rover or the sudden drop in power in solar PV power plants can be explained by the complex interaction of radiation data, wind speed and solar panel temperature. Existing rooftop PV plant data is a good laboratory resource to explore these complex events.